# FAST PHYLOGENY RECONSTRUCTION THROUGH LEARNING OF ANCESTRAL SEQUENCES

RADU MIHAESCU, CAMERON HILL, AND SATISH RAO

ABSTRACT. Given natural limitations on the length DNA sequences, designing phylogenetic reconstruction methods which are reliable under limited information is a crucial endeavor. There have been two approaches to this problem: reconstructing partial but reliable information about the tree ([18, 7, 5, 13]), and reaching "deeper" in the tree through reconstruction of ancestral sequences. In the latter category, [6] settled an important conjecture of M.Steel, showing that, under the CFN model of evolution, all trees on $n$ leaves with edge lengths bounded by the Ising model phase transition can be recovered with high probability from genomes of length $O(\log n)$ with a polynomial time algorithm. Their methods had a running time of $O(n^{10})$.

Here we enhance our methods from [5] with the learning of ancestral sequences and provide an algorithm for reconstructing a sub-forest of the tree which is reliable given available data, without requiring a-priori known bounds on the edge lengths of the tree. Our methods are based on an intuitive minimum spanning tree approach and run in $O(n^3)$ time. For the case of full reconstruction of trees with edges under the phase transition, we maintain the same sequence length requirements as [6], despite the considerably faster running time.

1. Introduction

Reconstructing the pattern of common ancestry among species is a central problem in evolutionary biology. This pattern is most commonly represented as a phylogenetic tree: a rooted tree with leaf-set $\delta(T)$ labeled by the species (or taxa) in $X$. Furthermore, it is generally assumed that phylogenies are binary: every speciation event is a divergence of two species from one common ancestor. Therefore the nodes $V(T)$ of the tree are either leaves corresponding to extant species in the set $X$, or internal nodes of degree three, corresponding to the ancestral species at each speciation event.

The phylogeny reconstruction problem is to discern the tree that accurately represents the evolutionary history of the taxa $X$. It is natural to identify each taxon with its genetic sequence and exploit molecular level differences between species to recover the phylogeny. To render the reconstruction problem tractable, it is commonly assumed that genetic sequences are correctly aligned and that sequences at the leaves are evolved from a root sequence according to an evolutionary Markov process on the tree: each edge $e$ in the tree, corresponding to an ancestral "divergence event", is equipped with mutation probability matrix $P(e)$. The sites of the sequences are evolved identically and independently according to these mutation probabilities.

The amount of disagreement between two sequences will then, depending on the underlying model of evolution, provide a scalar distance measure between the two sequences. As we detail in the next section, under suitable independence assumptions these distances are additive: the distances between the leaf taxa $X$ will correspond to the graph distance $D_T$ given by edge lengths $L$ on the tree $T$.

Most phylogeny reconstruction algorithms rely on estimating pairwise distances between taxa from the available genetic sequences and, in turn, using these estimates to recover topological information. Intuitively, reconstruction is achieved by piecing together topologies of smaller sub-trees which have a uniquely defined supertree. For instance, it is a fundamental result that a binary phylogenetic tree can be correctly recovered from its quartets: topologies describing the ancestral relations between subsets $X' \subset X$, $|X'| = 4$.

The main difficulty in the reconstruction of full phylogenies lies in the correct identification of short and deep divergence events [12, 8]. Intuitively, a divergence event is correctly recovered when the amount of mutation it induces is not drowned by mutation along the evolutionary paths leading away from it. Like any statistical estimator, the accuracy of evolutionary distance estimates is increasing in the amount of available data (length of the genetic sequences), but naturally decays as the variance in the system grows. In our case, longer biological distances have higher variance and are harder to estimate correctly. The probability of correctly resolving an ancestral divergence event is therefore naturally decreasing in the length of the pairwise distances used in its discovery, and increasing in the length of its corresponding edge.

It has been shown previously that, given upper and lower bounds on the mutation probabilities along each edge of $T$, $N = \log^{O(1)} n$ sites will suffice to reconstruct $T$ correctly for almost all topologies $T$. Intuitively, this approach relies on the fact that most phylogenies are not very deep: all internal nodes $v$ have enough descendants among the observable present species that are a bounded number of edges away, and whose observed character sequences thus provide enough information to resolve the topological structure of $T$ around $v$.

However, for topologies containing very deep nodes (such as in perfectly balanced trees), the reconstruction requires accurate estimation of distances between taxa that are "far-apart", therefore necessitating longer character sequences. Indeed, [16] shows that in the case of perfectly balanced binary trees, $N = n^{O(1)}$ is required for accurate reconstruction.

Recently it has been a growing trend to design algorithms that do not always attempt to recover a full tree ([18],[7], [13] and our own [5]), but only provide topological information that can be reliably extracted from the data, generally in the form of a forest of edge-disjoint subtrees of the original tree. This is a very important feature of reconstruction algorithms, as most real data-sets are not sufficient for recovering a full topology, and therefore any algorithm designed to return a full tree is bound to also give possibly incorrect information.

Another possible source of improvement in the area involves the reconstruction of internal genomes, which therefore provides pairwise distance estimates between internal nodes, allowing us to reach "deeper" in the topology and reconstruct from shorter distances. This method was introduced by Mossel [17] for the CFN model of evolution. He showed that for any fixed topology on $n$ leaves with edge lengths less than $\lambda_0 = \log(2)/4$, the so-called "phase transition of the Ising model on trees", arbitrarily deep internal sequences can be recovered with bounded probability of error. This implies that leaf sequences of length $O(\log n)$ suffice ro distinguish between all perfectly balanced phylogenies on $n$ leaves. A simple information-theoretic argument shows that this bound is tight.

Mossel's techniques were then used by [6] in the context of phylogeny reconstruction. Given a lower bound $f$ and an upper bound $g < \lambda_0$ for the edge lengths of $T$, [6] show that the full topology can be recovered from sequences of length $O(\log n)$, thereby settling an important conjecture of M.Steel. Their algorithm has a worst case running time of $O(n^{10})$. In [8] it is shown that $N$ must grow at least as fast as $O(\log n)$, and therefore the results of [6] are asymptotically optimal. The results of [6] have been partially extended by Roch [19] to a general time-reversible model, with worse but still sub-polynomial sequence length requirements.

Here, we will present a relatively simple algorithm which combines our approach in [5] with the reconstruction of ancestral sequences as detailed in [17]. The ability to learn ancestral sequences is central to the success of the methods in [6] and subsequently our methods, as it allows the reconstruction of the model tree topology $T$ by piecing together quartet topologies of bounded diameter on its internal nodes. By contrast, previous algorithms that have only looked at quartets on the leaves of $T$ have achieved strictly weaker results.

For trees with edges under the phase transition we achieve full topology recontruction with the same edge length requirements as [6]. Our algorithm relies on an intuitive minimum spanning tree approach: we progress recursively by growing an edge-disjoint sub-forest of $T$. The algorithm halts when no further progress can be made reliably, meaning that all edges that could be added are either too short to be resolved accurately, or they violate the phase transition bound, therefore preventing further reliable reconstruction of ancestral genomes.

Our contributions here are threefold:
- We reduce the worst case running time to $O(n^3)$, thus matching that of much simpler phylogeny reconstruction algorithms, such as Neighbor-Joining.
- In the case when full reconstruction is not possible with the available data, we return reliable partial information in the form of an edge-disjoint sub-forest of $T$.
- We eliminate the need for a-priori knowledge of the edge length bounds $f$ and $g$. Rather, we infer an edge length tolerance interval from the length of the available genetic sequences and reconstruct pieces of the tree with edges within this interval.

It is worth noting that our edge length tolerance interval can in fact be controlled by the user. Increasing it can potentially result in a larger output forest, but will trade off against the expected accuracy of this output. We also note that our method implies similar results for all group based models of evolution where the character alphabet is a group $G$ admitting a non-trivial morphism $\phi : G \to \mathbb{Z}_2$. This class of models includes, among others, the well known Kimura 3ST [15] and Jukes-Cantor models. We elaborate on this technical point in Appendix B.

## 2. Background on phylogeny reconstruction

In this work, we concentrate on the Cavender-Farris-Neyman (CFN) 2-state model of evolution ([3],[10]): our genetic sequences are bit strings of some length $N$ and the probability of mutation $p(e)$ along an edge $e$ of the tree does not depend on the starting state. We denote the $i$'th entry of the sequence corresponding to taxon $a \in X$ as $\chi_i(a)$. The vectors $\chi_i(:)$ are also known as characters of the set $X$.

For each position $i$, the character values at the nodes of $T$ mutate independently along each edge $e = (u, v) \in E(T)$, starting from a uniform distribution at the root node $\rho$, according to the symmetric transition matrix

$$M(e) = \exp(L(e)R) = \begin{pmatrix} 1 - p(e) & p(e) \\ p(e) & 1 - p(e) \end{pmatrix},$$

where $R$ is the symmetric rate matrix $\begin{pmatrix} 1 & -1 \\ -1 & 1 \end{pmatrix}$. Then $L(e) = -\log(1 - 2p(e))/2$, $p(e) = \mathbb{P}(\chi_i(u) \neq \chi_i(v))$ and the distribution of character states at any node is also uniform.

The topology $T$, together with the edge lengths $L$ define a joint probability distribution $\mathbb{P}'_{T,L}$ on the character values at the nodes of $T$ and $\chi_i(:)$ are i.i.d. samples from this probability distribution. Note that the values of $\chi_i(u)$ are not known for ancestral nodes $u \in V(T) \setminus X$. We therefore define $\mathbb{P}_{T,L}$ to be the marginal distribution of $\mathbb{P}'_{T,L}$ at the leaves $X$. The *observed* character sequences $\chi_i(X)$ are then i.i.d. samples from $\mathbb{P}_{T,L}$.

Let $\Omega$ be the set of all possible binary topologies on $X$. The problem of phylogeny reconstruction is then equivalent to finding an algorithm, or estimator, $A : \{\pm 1\}^{|X| \times N} \longrightarrow \Omega$, such that the probability that $A(\chi_1, \ldots, \chi_N) = T$ is maximized. As with most estimation problems, the central question then becomes: how many samples $N$ do we need in order to achieve accurate reconstruction of the underlying tree?

**Note:** In the case of the CFN model we can only recover the "un-rooted" topology $T$, but not the location of the earliest specie in $T$, i.e. its root. This is because CFN is a reversible model of evolution, meaning that the probability distribution $\mathbb{P}'_{T,L}$ does not depend on the location of the root $\rho$, and therefore neither does $\mathbb{P}_{T,L}$. See [20] for more details on Markov models of evolution.

For two uniform Bernoulli variables $u, v$, sharing a joint distribution $\mathbb{P}$, let us define $D_{\mathbb{P}}(u, v) = -\log(1 - 2\mathbb{P}[u \neq v])/2$. Note the similarity to the definition of the edge lengths under the CFN model. It is easy to check that for three uniform Bernoulli variables $v_i$, with $i \in \{1, 2, 3\}$, such that $(v_1 \perp\!\!\!\perp v_3 | v_2)$, the following holds:

$$D(v_1, v_3) = D(v_1, v_2) + D(v_2, v_3).$$

Here $(v_1 \perp\!\!\!\perp v_3 | v_2)$ means that $v_1$ and $v_3$ are independent conditioned on the value of $v_2$. In other words, given the Markov property of the CFN model (see [20]), for two nodes $a, b \in V(X)$ joined by a path $p$, we have the following relationship:

$$D(\chi(a), \chi(b)) = \sum_{e \in p} L(e).$$

Here and in the remainder of our paper, $D$ is the distance given by the joint probability distribution $\mathbb{P}'_{T,L}$.

This implies that knowing the joint probability distribution of character values at pairs of leaves will provide us with the distance between the two leaves according to the edge lengths defined above, which will in turn provide the topology $T$ and the individual edge lengths $L$. In practice we will, of course, not know $D$ precisely, but we will be able to estimate it from the observed character values $\chi_i(:)$, which are i.i.d. samples from the marginal $\mathbb{P}_{T,L}$. For $a, b \in X$, define

$$\hat{D}(a, b) = -0.5 \log(1 - \frac{2}{N} \sum_i \mathbf{1}[\chi_i(u) \neq \chi_i(v)]) = -0.5 \log(\frac{1}{N} \sum_i \chi_i(u)\chi_i(v)).$$

Consider the simplest example of reconstructing a *quartet*: a binary topology $Q$ on 4 leaves $X = \{a, b, c, d\}$ (there is only one possible topology on 3 leaves). There are three possibilities, each corresponding to a pairing of the four taxa. We let $Q = (a, b|c, d)$ encode the case when the taxa $a, b$ are separated from the taxa $c, d$ by an edge $e$. In the case $Q$ is indeed the correct topology on the four taxa, the true pairwise distance matrix $D$ satisfies the so-called *four point condition*:

$$D(a, b) + D(c, d) < D(a, c) + D(b, d) = D(b, c) + D(a, d),$$

and moreover $2L(e) = D(a, c) + D(b, d) - D(a, b) + D(c, d)$, where $D(a, b) = D(\chi(a), \chi(b))$ for ease of notation.

Given the approximate distance $\hat{D}$, the procedure FPM, as in *four point method*, gives us a way to resolve the topology of the quartet $a, b, c, d$, while the procedure ME, as in *middle edge*, gives us a way to estimate internal edge lengths. We borrow some of our notation from [9].

**Definition 2.1.** *Suppose $\hat{D}(a,b)+\hat{D}(c,d) < \hat{D}(a,c)+\hat{D}(b,d) \leq \hat{D}(b,c)+\hat{D}(a,d)$. Then let*

$$FPM(\hat{D}; a, b, c, d) = (a, b|c, d) \text{ and}$$
$$ME(\hat{D}; (a, b|c, d)) = (\hat{D}(a,c) + \hat{D}(b,d) + \hat{D}(b,c) + \hat{D}(a,d) - 2\hat{D}(a,b) - 2\hat{D}(c,d))/4.$$

We observe that as long as $|\hat{D}(i,j) - D(i,j)| < \epsilon/2 < L(e)/2$, for $i, j \in \{a, b, c, d\}$, then FPM recovers the correct quartet topology, and that $|ME(\hat{D}, Q) - L(e)| < \epsilon$, where $Q = (a, b|c, d)$.

It is a fundamental fact in phylogenetics that the topology of the entire tree can be recovered from the topologies of its quartets (see [20] for details). The following proposition is the first step towards giving lower bounds on the number of samples $N$ that insure proper reconstruction. Its proof is implied by the proof of Theorem 8 in [9] and has been proved in several other publications.

**Theorem 2.2.** [9] *Let $u, v$ be uniform binary random variables with $\mathbb{P}(u \neq v) < y$. Given $N$ samples of $u, v$ and the associated empirical distance $\hat{D}$, then*

$$\mathbb{P}[\hat{D}(u,v) > D(u,v) + \epsilon/2] < 1.5 \exp\left[\frac{-(1-\sqrt{1-2z})^2(1-2y)^2 N}{8}\right],$$
$$\mathbb{P}[\hat{D}(u,v) < D(u,v) - \epsilon/2] < 1.5 \exp\left[\frac{-(1-\sqrt{1-2z})^2(1-2y)^2 N}{8}\right],$$

*where $\epsilon = -\log(1-2z)/2$.*

Theorem 2.2 has the following easy but important corollary: in a nutshell, given a fixed $y$ and $M = -\log(1-2y)/2$, distances larger than $M$ will, with high probability, be "estimated" as longer than $M - \epsilon$. The proof comes from the second inequality of Theorem 2.2 via a standard coupling argument, and is therefore omitted.

**Corollary 2.3.** *Let $u, v$ be uniform binary random variables with $\mathbb{P}(u \neq v) > y$. Given $N$ samples of $u, v$ and the associated empirical distance $\hat{D}$, then*

$$\mathbb{P}[\hat{D}(u,v) < M - \epsilon/2] < 1.5 \exp\left[\frac{-(1-\sqrt{1-2z})^2(1-2y)^2 N}{8}\right],$$

*where $\epsilon = -\log(1-2z)/2$ and $M = -\log(1-2y)/2$.*

In general, when an estimator $\hat{D}$ of the quantity $D$ satisfies

$$|\hat{D} - D| < \epsilon/2 \text{ when } D < M$$
$$\hat{D} > M - \epsilon/2 \text{ when } D > M$$

we say that $\hat{D}$ is an $(M, \epsilon/2)$-*estimator* for $D$. This is a very slight modification of the concept of $(\epsilon, M)$-distortion from [18].

Suppose $g > L(e) > f, \forall e \in E(T)$, with $f, g > 0$ fixed. Let $f > \epsilon = -\log(1 - 2z)$ and $M = -\log(1 - 2y)/2$. Let $A$ be an algorithm which attempts to recover the full topology of $T$ by evaluating $K = O(n^k)$ empirical distances between pairs of random variables $u, v$. Suppose in addition that $A$ recovers the correct topology if ALL the emprical distances inspected are $(M, \epsilon/2)$-approximations of the true distances. Theorem 2.2 and Corollary 2.3 guarantee that the empirical distance matrix $\hat{D}$ satisfies this property, with high probability, provided the number of samples $N$ is large enough. If we want to ensure that $\mathbb{P}[A(\hat{D}) \neq T] < 1 - p$ for some $p > 0$, plugging into the above inequality yields

$$N \geq O(e^{4M} k \log n).$$

This inequality is essential to understanding the need for learning ancestral sequences. Indeed, given that the topological depth of an internal node can grow as high as $O(\log n)$, any method which is restricted to inspecting pairwise distances between leaves of $T$ will have to estimate distances as high as $M = O(g \log n)$, which yields $N = n^{O(1)}$. By contrast, learning ancestral sequences gives us a way to resolve the entire topology by only inspecting $K = O(n^2)$ distances between nodes separated by at most a constant distance. If the edge lengths are under the phase transition, as explained in the following section, we can guarantee that the additional noise coming from estimating internal sequences is also bounded by a fixed amount. Thus $M = O(1)$ and thus $N = O(\log n)$.

## 3. Background on learning ancestral characters

In this section we will show how to recover the sequences at the interior nodes of a phylogenetic tree from the sequences at the leaves of the tree, up to an *a priori* bounded error. We do this by means of a recursive majority algorithm. All the results in this section have appeared in previous publications, such as [17], and are used in an identical manner in [6]. For this reason we will state them without proof.

**Definition 3.1.** *Given a sequence of $\pm 1$ bits $x_1, \ldots x_n$, we define the majority function*

$$Maj(x_1, \ldots x_n) = sign(x_1 + \ldots + x_n + .5w),$$

*where $w$ is an unbiased $\pm 1$ random variable that is independent of the $x_i$'s.*

**Definition 3.2** (Definition 4.1 in [17]). *Let $T = (V, E)$ be a tree rooted at $\rho$ with leaf-set $\delta T$. For functions $l : E \to [0, \infty]$ and $\eta : \delta T \to [0, \infty]$, let $CFN(l, \eta)$ be the CFN model on $T$ where*

- *the edge length $L(e)$ is equal to $l(e)$ for all $e \in E$ not adjacent to $\delta T$*
- *$L(e) = l(e) + \eta(v)$ for all edges $e = (u, v)$ with $v \in \delta T$.*

*Let $\widehat{Maj}(l, \eta) = D(\chi(\rho), Maj(\bar{\chi}(\delta T)))$, where $\bar{\chi}$ are the character values on $T$ given by $CFN(l, \eta)$.*

The distance $D$ is the one provided by the joint probability distribution defined by $CFN(l, \eta)$, together with the independent coin-flips necessary for breaking ties in the Maj function. In other words, $\widehat{Maj}(l, \eta) = -\log(1 - 2\mathbb{P}[\chi(\rho) \neq \text{Maj}(\chi(\delta T))])/2$.

The intuition behind the above definition is as follows. Let $\chi$ be the character values at nodes of $T$, defined according to the CFN model given by edge lengths $l$ on $T$. Suppose that the character value $\chi(u)$ at each leaf $u \in \delta(T)$ is perturbed by an independent noise source such that the probability of perturbation is $(1 - \exp(-2\eta(u)))/2$. Let $\tilde{\chi}(u)$ be the perturbed character value, so formally:

$$\tilde{\chi}(u) \perp\!\!\!\perp \chi(v) | \chi(u), \forall v \in V(T) \text{ and } \tilde{\chi}(u) \perp\!\!\!\perp \tilde{\chi}(v) | \chi(u), \forall v \in \delta(T) \tag{1}$$

$$\mathbb{P}[\tilde{\chi}(u) \neq \chi(u)] = (1 - \exp(-2\eta(u)))/2 \Leftrightarrow D(\tilde{\chi}(u), \chi(u)) = \eta(u). \tag{2}$$

It is then an easy exercise to verify that our definition of $\bar{\chi}$ is equivalent to the one below:

$$\bar{\chi}(u) = \begin{cases} \chi(u) \text{ for } u \notin \delta(T) \\ \tilde{\chi}(u) \text{ for } u \in \delta(T). \end{cases}$$

For our purposes, the noise at the leaves of the subtree arises from the reconstruction of the character values by way of recursive majority. Our hope is that we can design a recursive learning procedure such that the probability of error $\mathbb{P}[\chi'(u) \neq \chi(u)]$ remains bounded away from .5 as we progress deeper and deeper into $T$. Theorem 3.3 achieves this remarkable feat. Our formulation of the theorem is a specialization of Theorem 4.1 in [17] to binary trees and we state it without proof.

**Theorem 3.3** (Theorem 4.1 in [17]). *Let*

$$a(q) = 2^{1-q} \lceil \frac{q}{2} \rceil \binom{q}{\lceil \frac{q}{2} \rceil}.$$

*For $d \in \mathbb{Z}_{>0}$, $\lambda_{max} > 0$ and $0 < \alpha(\lambda_{max}) < a(2^d)e^{-2d\lambda_{max}}$, there exists $\beta(\lambda_{max}) > 0$, such that the following hold. Let $T$ be a $d$-level balanced binary tree and consider the $CFN(l, \eta)$ model on $T$, where $\max l \leq \lambda_{max}$ and $\max \eta \leq \eta_{max}$. Then*

$$\widehat{Maj}(l, \eta) \leq \max\{\eta_{max} - \log(\alpha)/2, \beta\}. \tag{3}$$

Using Stirling's approximation formula, it can be shown that $a(q) \approx \sqrt{\frac{2}{\pi}}\sqrt{q}$. For $\lambda_{max} = \lambda_0 - \epsilon$ with $\lambda_0 = \log(2)/4$ and $\epsilon > 0$ (i.e. under the phase transition), we have

$$a(2^d)e^{-2d\lambda_{max}} > a(2^d)e^{-2d(\lambda_0 - \epsilon/2)} \approx \sqrt{\frac{2}{\pi}}e^{\epsilon d},$$

thus for $d$ large enough $a(2^d)e^{-2d\lambda_{max}} > 1$. Setting $\alpha = 1$ and $\eta_{max} = \beta$ in Theorem 3.3 we obtain the following corollary:

**Corollary 3.4.** *For $0 < \lambda_{max} < \lambda_0$, there exists $d_0 > 0$ such that: for any $d > d_0$, there exists $\beta(\lambda_{max}, d) < \infty$, such that for any balanced $d$-level binary tree $T$ and any functions $l : E(T) \to [0, \lambda_{max}]$, $\eta : \delta T \to [0, \beta]$, we have $\widehat{Maj}(l, \eta) \leq \beta$.*

To put Corollary 3.4 in words, for trees with edge lengths less than $\lambda_{max}$, learning ancestral character sequences via recursive majority on sub-trees of height $d$, as detailed below, gives learned character sequences whose distance to the true sequences is recursively bounded by $\beta$. This is the crucial result for the development of our algorithm, as it implies that recursive reconstruction with reliable, non-decaying accuracy is possible on trees of any size.

Indeed, let us suppose that $l(e) < \lambda_{max} = \lambda_0 - \epsilon$ for all $e \in E(T)$. Let $d > d_0$ and $\beta$ be as in the above corollary. We decompose $T$ recursively into a collection of edge disjoint rooted trees in the following manner: start from the root $\rho$ and follow all paths down the tree until each path reaches length $d$ or terminates with a leaf. Cut the tree at the endpoint of each path and recurse on the subtrees rooted at these endpoints. This procedure divides $T$ into trees of depth at most $d$. Let $T_1, \ldots T_k$ be the collection of trees in the subdivision of $T$ and let $\rho_i$ be the root of $T_i$ for all $i$. See Figure 3(a).

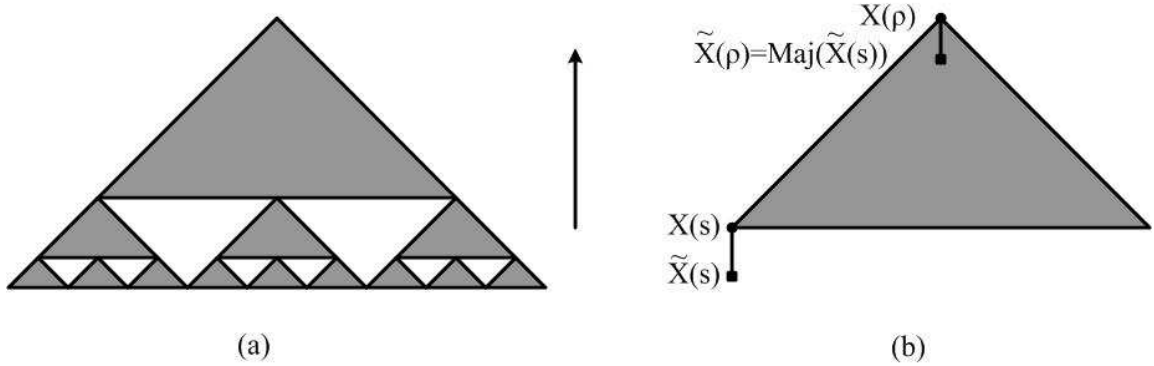

FIGURE 1. Learning ancestral sequences by bottom-up recursion.

We can now define a recursive learning process. The learned character value $\tilde{\chi}(v)$ is set equal to $\chi(v)$ for all $v \in \delta T$. For each subtree $T_i$ such that the value $\tilde{\chi}(v)$ has been specified for all $v \in \delta T_i$, we define $\tilde{\chi}(\rho_i) = \text{Maj}(\tilde{\chi}(\delta T_i))$. Now recurse as in Figure 3(b).

Note that some of the subtrees $T_i$ may not be fully balanced as required by Theorem 3.4. Suppose $u \in \delta T_i$ and the topological distance between $u$ and $\rho_i$ is $k < d$. In this case we replace $u$ by a balanced binary tree of height $d - k$ with all edges of length 0, which is equivalent to giving $\chi(u)$ weight $2^{d-k}$ in the $\text{Maj}(\tilde{\chi}(\delta T_i))$ vote. For clarity of exposition, we will keep the notation Maj to represent this weighted majority.

**Theorem 3.5.** *Suppose* $\max\{l(e) : e \in E(T)\} = \lambda_{max} < \lambda_0 - \epsilon$ *with* $\epsilon > 0$, *and let* $d$ *and* $\beta$ *be as in Corollary 3.4. The procedure described above gives a bottom up learning process which ensures that* $D(\tilde{\chi}(\rho), \chi(\rho)) < \beta$.

**Proof of Theorem 3.5:** Set $\eta(u) = D(\chi(u), \tilde{\chi}(u))$ for all $u \in \delta T$ or $u = \rho_i$ for some $i$. We prove the following two conditions by bottom-up induction on the sub-trees $T_i$:
- $\tilde{\chi}(u) \perp\!\!\!\perp \chi(v) | \chi(u), \forall v \in V(T_i), u \in \delta T_i$ and $\tilde{\chi}(u) \perp\!\!\!\perp \tilde{\chi}(v) | \chi(u), \forall v \in \delta(T)$
- $\eta(\rho_i) < \beta$.

First, for all $u \in \delta T$, $\tilde{\chi}(u) = \chi(u)$, so $\eta(u) = 0$. Both hypotheses are thus obeyed trivially for subtrees formed by a single leaf. This provides the base case for our induction.

Now consider a subtree $T_i$ and suppose $\eta(u) < \beta$ for all $u \in \delta T_i$. If $u \in \delta T$, the first induction hypothesis is obeyed trivially, as $\tilde{\chi}(u) = \chi(u)$. Alternatively, suppose $u = \rho_j$. Let $\bar{T}$ be the subtree of $T$ rooted at $u$. Then $\tilde{\chi}(u)$ is a function of the values $\chi(\delta \bar{T})$ and moreover the Markov property of the CFN model implies that $\chi(\delta \bar{T}) \perp\!\!\!\perp \chi(v) | \chi(u)$, for all $v \in V(T_i)$. Therefore $\tilde{\chi}(u) \perp\!\!\!\perp \chi(v) | \chi(u)$. The other statement of the first induction hypothesis follows similarly.

Finally, Corollary 3.4 implies
$$\eta(\rho_i) = D(\chi(\rho_i), \tilde{\chi}(\rho_i)) = D(\chi(\rho_i), \mathrm{Maj}(\tilde{\chi}(\delta T_i))) = \widehat{Maj}(l(E_{T_i}), \eta_{\delta T_i}) \leq \beta,$$
so the second induction hypothesis is also obeyed. □

## 4. General outline of the algorithm TREE-MERGE

Let $\lambda_0$ be the phase transition. Suppose the set of taxa $X$ has cardinality $n$ and the character sequences identifying the taxa have length $N$. Given $\epsilon > 0$ we define the following quantities:
- $\lambda_{max}(\epsilon) = \lambda_0 - \epsilon$
- $d(\lambda_{max}(\epsilon))$ and $\beta(\lambda_{max}(\epsilon))$ are the depth of the trees in the recursive majority decomposition and the upper bound on the learning noise, as in Corollary 3.4.
- $M(\epsilon) = 24\lambda_0 + 6\beta(\lambda_{max}(\epsilon)) + 12\epsilon$.

Given the length $N$ of available sequences, the number of taxa $n$, and a user-define maximum allowed probability of error $\xi$, we can pick $\epsilon$ such that
$$1.5 \exp\left[\frac{-(1-e^{-\epsilon})^2 e^{-4M} N}{8}\right] < \frac{\xi}{16n^2}. \tag{4}$$

By Theorem 2.2 and Corollary 2.3, for any two character values $u, v$, learned or observed, drawn from the joint probability distribution $\mathbb{P}'_{T,L}$, the empirical distance $\hat{D}(u,v)$ will be an $(M, \epsilon/2)$-approximation for the true distance $D(u,v)$, with probability at least $1 - n^{-2}\xi/8$. When an event happens with probability at least $1 - O(n^{-2}\xi)$, we say that it occurs *with high probability*. By Lemma 4.2, TREE-MERGE will evaluate no more than $8n^2$ empirical distances. By the union bound with probability at least $1 - \xi$, the following condition holds:

**Condition 4.1** ($\star$). *All the empirical distances evaluated by our algorithm are $(M, \epsilon/2)$-approximations of the corresponding true distances.*

**Lemma 4.2.** *Algorithm TREE-MERGE reconstructs at most $3n$ ancestral sequences in addition to the $n$ sequences at the leaves, and thus computes at most $8n^2$ pairwise distances.*

**Corollary 4.3.** *By the union bound applied to equation (4), with probability $1 - \xi$ all the empirical distances observed by TREE-MERGE are $(M, \epsilon/2)$-approximations of the corresponding true distances. Thus condition ($\star$) holds with probability $1 - \xi$.*

**Proof of Lemma 4.2:** Indeed, there are $n-3$ internal nodes in any binary tree on $n$ leaves. Any internal vertex of any subtree is also a node in the parent tree. Our algorithm progresses by joining at each step a pair of components of the forest through the addition of a new edge, creating zero, one or two more nodes of the forest. No nodes are ever destroyed.

We define a *clade* of a tree $T'$ to denote a subtree of $T'$ that is induced by removing an edge $e \in E(T)$. Each edge $e$ defines two clades and for each clade there is a natural rooting at the corresponding endpoint of $e$. For each internal node $v$ of a forest, there are three clades rooted at $v$, each one induced by one of the edges adjacent to $v$. These clades correspond to the three "directions" leading away from $v$.

Inspection of the algorithm TREE-MERGE shows that an internal sequence corresponding to a node/direction pair is learned by TREE-MERGE when the corresponding clade becomes "proper" (see Section 6 for definition). Once a sequence is learned, it gets stored and is never modified, regardless of new growth in the corresponding clade. Each internal node of the tree will have exactly three learned sequences, each being constructed exactly once. Thus TREE-MERGE inspects at most $n+3(n-2)$ sequences and at most $8n^2$ sequence pairs. □

We will prove that under condition $(\star)$, TREE-MERGE will recover a topologically correct forest of edge-disjoint subtrees of the model tree $T$. If, in addition, the conditions of Theorem 4.5 hold, then TREE-MERGE will recover the entire tree. In the subsequent treatment we will generally assume that condition $(\star)$ holds, unless otherwise stated.

The algorithm TREE-MERGE progresses, as the name suggests, by gradually building a sub-forest $F$ of $T$, such that the following three invariants are obeyed:

**I1**: For any component $T_i \in F$ and any edge $e \in E(T_i)$, the path corresponding to $e$ in $T$ has length at least $2\epsilon$.

**I2**: For any component $T_i \in F$, all edges of $T_i$ *except at most one* have corresponding paths in $T$ of length at most $\lambda_0 - \epsilon$, and all edges have corresponding paths of length at most $2\lambda_0 - 4\epsilon$.

**I3**: Any two connected components $T_i, T_j \in F$ are edge disjoint as subgraphs of $T$.

Invariant **I1** is needed in order to ensure that reconstructed ancestral divergence events are long enough to be reliable. **I2** guarantees that ancestral sequences can be learned reliably from subtrees with edges under the phase transition. Finally **I3** is a technical requirement of the algorithm. It allows us to reliably resolve topological information despite conditional dependencies between learned ancestral sequences.

In order to ensure that **I1** and **I2** hold however, we need a reliable way to estimate edge lengths. As observed in Section 2, $(\star)$ guarantees that edge lengths which can be estimated as middle paths of quartets with diameter less than $M$ will have an estimation error less than $\epsilon$. In Lemma 6.2 we show formally that TREE-MERGE will in fact estimate all edge lengths within $\epsilon$ error.

Given this fact, we can now enforce **I1** and **I2** by requiring instead that the following two conditions hold:

---

Algorithm $F$ = TREE-MERGE$(X, \chi)$
INPUT: n binary sequences $\chi$ of length N corresponding to taxa $X$.
OUTPUT: An unrooted forest $F$ detailing partial information on the evolutionary relationships of the taxa $X$.
  (1) set $F = X$, i.e. $F$ contains trees formed by single nodes.
  (2) insert all leaf taxa distances in NodeDistList.
  (3) insert all tree distances less than $M/3 - \epsilon$ in TreeDistQueue.
  (4) **while** $|F| > 1$
      (a) **if** TreeDistQueue $= \emptyset$, **return** $F$
      (b) $(T_1, T_2)$ = pop(TreeDistQueue). Let $(E_1, E_2)$ = TreeConnection$(T_1, T_2)$
      (c) **if** $|E_1| > 1$ **or** $|E_2| > 1$ **continue**
      (d) Let $Q = (a, b : c, d)$, where $E_1 = \{(a,b)\}$, $E_2 = \{(c,d)\}$. Set $T_{new} = T_1 \cup T_2 \cup Q$ and compute the edge lengths of the quartet $Q$.
      (e) **if** $T_{new}$ violates condition **C1**, **continue**
      (f) **if** $T_{new}$ violates condition **C2 continue**
      (g) **if** $\exists T_k$ s.t. TreeDistance$(T_1, T_2) + 3\epsilon >$ TreeDistance$(T_1, T_k)$ + TreeDistance$(T_k, T_2)$, **continue**
      (h) **else**
            (i) $F = F \setminus \{T_1, T_2\} \cup \{T_{new}\}$.
            (ii) compute learned characters for all new roots of proper clades of $T_{new}$.
            (iii) insert all distances involving new learned characters in NodeDistList.
            (iv) UpdateTreeDistQueue$(T_1, T_2, T_{new})$
  (5) **return** F

FIGURE 2. Algorithm TREE-MERGE.

  **C1**: Each edge in $F$ has estimated distance at least $3\epsilon$.
  **C2**: For each $T_i \in F$, the edges of $T_i$ have estimated length at most $\lambda_0 - 2\epsilon$, with the exception of at most one edge, whose estimated length is less than $2\lambda_0 - 5\epsilon$. We call such an edge a *long* edge.

Philosophically, our approach is very similarly to the classical minimum spanning tree algorithms. At each step of the algorithm we will join two connected components $T_i, T_j$ such that the new component does not violate **C1** and **C2**, and the estimated length of the path linking $T_i$ and $T_j$ is the shortest among all candidate pairs. This by itself does not guarantee **I3**. However, condition $(\star)$ and step 4.(g) of TREE-MERGE achieve this purpose, as will be shown in formally in Section 6.

We can now state the three main results of this paper. We postpone the formal proofs until Section 6. All of our results assume that $N, \xi$ and $\epsilon$ are such that (4) holds, which in turn guarantees that condition $(\star)$ holds with probability at least $1 - \xi$.

**Theorem 4.4.** *If $(\star)$ holds, algorithm TREE-MERGE returns a topologically correct sub-forest $F$ of $T$ satisfying invariants **I1**, **I2**, **I3**.*

---

Algorithm UpdateTreeDistQueue($T_1, T_2, Q$)
INPUT: Subtrees $T_1$ and $T_2$ to be joined and old TreeDistQueue data-structure containing distances between "communicating" pairs of trees in $F$.
OUTPUT: Updated TreeDistQueue data structure.
**for** $k \neq i, j$
   (1) Let $E_1 = \{(a, b)\}$ and $E_2 = \{(c, d)\}$, where $Q = (a, b : c, d)$.
   (2) **if** $(T_1, T_k)$ or $(T_2, T_k)$ were ever in TreeDistQueue
       • For $i = 1, 2$, let $(E'_i, E_k) = $ TreeConnection$(T_i, T_k)$ if $(T_i, T_k) \in$ TreeDistQueue, or $E'_i = E(T_i)$ otherwise.
       • If $E_k$ has not been set, let $E_k = E(T_k)$.
       • **if** $E'_1 \neq E_1$ set $E_{new} = E'_1$.
       • **elseif** $E'_2 \neq E_2$ set $E_{new} = E'_2$.
       • **else** Set $E_{new} = E(Q)$
       • $(E_{new}, E_k) = $ TreeConnection$(T_{new}, T_k, E_{new}, E_k)$.
   (3) **elseif** $\tilde{D}(\tilde{u}, \tilde{t}) < M/3 - \epsilon$ or $\tilde{D}(\tilde{v}, \tilde{t}) < M/3 - \epsilon$ for some $t \in V(T_k)$, then
       • $(E_{new}, E_k) = $ TreeConnection$(T_{new}, T_k, E(T_{new}), E(T_k))$.
   (4) **if** a connection was found above
       • $d = $ TreeDistance$(T_{new}, T_k, E_{new}, E_k)$.
       • TreeDistQueue $= $ remove(TreeDistQueue, $(T_i, T_k), (T_i, T_k)$),
         TreeDistQueue $= $ insert(TreeDistQueue, $(T_{new}, T_k)$).
**end**

FIGURE 3. Subroutine UpdateTreeDistQueue updates the connections between trees in the forest $F$, after two components are merged.

**Theorem 4.5.** *Let $T$ satisfy $6\epsilon \leq L(e) \leq \lambda_0 - 3\epsilon, \forall e \in E(T)$. Then given $N$ independent samples $\chi_1, \ldots, \chi_N$ from the character distribution $\mathbb{P}_{T,L}$, $T$ will be fully and correctly recovered by TREE-MERGE with probability at least $1 - \xi$.*

**Theorem 4.6.** *TREE-MERGE always terminates in $O(Nn^2 + n^3)$ time, where the proportionality constant is a decreasing function of $\xi$ and $\epsilon$.*

Theorem 4.5 and equation (4) provide us with specific edge-length bounds for trees that can be reconstructed with probability at most $\xi$ from sequences of length $N$. Indeed, for any $N$ and $\xi$, there is a lower bound $\epsilon_{N,\xi}$ such that any $\epsilon > \epsilon_{N,\xi}$ satisfies inequality (4), which insures ($\star$) will hold.

This is an important feature of TREE-MERGE, as it allows us to recover an edge-length reliability interval from the available sequence lengths. In contrast, previous research has focused on recovering the necessary sequence lengths for full reconstruction, assuming that lower and upper bounds on edge lengths were known. As these bounds cannot be known a-priori, this is hardly useful, especially for algorithms which do not provide partial information in case full reconstruction is not feasible. We note however that under this paradigm, our methods still achieve the assymptotically best known sequence length requirements.

**Corollary 4.7.** *For trees $T$ satisfying $6\epsilon \leq L(e) \leq \lambda_0 - 3\epsilon$ for some fixed $\epsilon$, equation (4) and Theorem 4.5 show that $N = O_{\epsilon,\xi}(\log(n))$ sites suffice for TREE-MERGE to reconstruct $T$ with probability at least $1 - \xi$.*

We finally note that the constant factor in our running time bounds depends on the desired maximum probability of failure $\xi$, and on $\epsilon$. If one is to consider these parameters as fixed, then the running time is indeed simply $O(n^3 + n^2 N)$. A higher $\epsilon$ value implies a faster running time and shorter sequence length requirements, but trades off against the size of the reconstructed forest. Similarly, a higher value for $\xi$ implies shorter sequence length requirements, but trades off against the accuracy of the reconstructed forest. The specific dependencies between these parameters however are very complicated and beyond the scope of this paper.

## 5. A conditional independence toolkit

In this section we present four lemmas which are the main workhorses of our algorithms. All the results presented here hold in general for trees with arbitrary edge lengths, as they are qualitative statements which do not depend on the accuracy of the learned character values. This section, together with the proof of Lemma 5.4 in Appendix A, provide a stand-alone toolkit of useful new results in this area.

To simplify notation, here and in the remainder of the paper, for a node $v \in V(T)$ we will use $v$ to also denote the character value $\chi(v)$, as the distinction will be clear from context. Let an *induced* subtree $T'$ of $T$ be a subtree such that $\delta(T') \subset \delta(T)$. For an induced subtree $T'$ rooted at $\rho$, we will denote by $\tilde{\rho}(T')$ the character value $\chi'(\rho)$ that is "learned" from $\chi(\delta T')$ by recursive majority on $T'$, as described in the previous section. We also denote by $V(T) \cap T'$ the vertices of $T$ that are either in $V(T')$ or lie on the paths of $T$ corresponding to the edges of $T'$. Finally, for two nodes $u, v \in V(T)$ we let $P(u, v)$ be the path connecting $u$ and $v$ in $T$.

We let $D$ denote the distance between uniform Bernoulli random variables defined in the Introduction, where the underlying joint probability distribution is the one given by the CFN model on $T$, $\mathbb{P}'_{T,L}$ and the random coin tosses involved in the recursive majority learning of ancestral characters.

The following two lemmas are present and used almost identically in [6]:

**Lemma 5.1.** *Let $T_1, T_2$ be edge-disjoint subtrees of $T$ rooted at $\rho_1$ and $\rho_2$, such that $\delta T_1, \delta T_2 \subset \delta T$. Let $v_1 \in V(T) \cap T_1$ and $v_2 \in V(T) \cap T_2$ be the endpoints of the path $P(v_1, v_2)$ joining $T_1$ to $T_2$ along the edges of $T$. Then*

$$(\tilde{\rho}_1(T_1) \perp\!\!\!\perp v | x) \text{ and } (\tilde{\rho}_1(T_1) \perp\!\!\!\perp \tilde{\rho}_2(T_2) | x),$$

*for any $v \in V(T) \cap T_2$ and $x \in P(v_1, v_2)$. See Figure 4(a).*

**Proof of Lemma 5.1:** The nodes of $T_1$ are separated from those of $T_2$ by $x$. By the Markov property of the CFN model, $(\delta T_1 \perp\!\!\!\perp v | x)$. Since $\tilde{\rho}_1(T_1)$ is a deterministic function of $\delta T_1$ and independent coin flips (tie-breakers in the recursive majority), we conclude $(\tilde{\rho}(T_1) \perp\!\!\!\perp v | x)$. The proof of the second statement is almost identical and is omitted. □

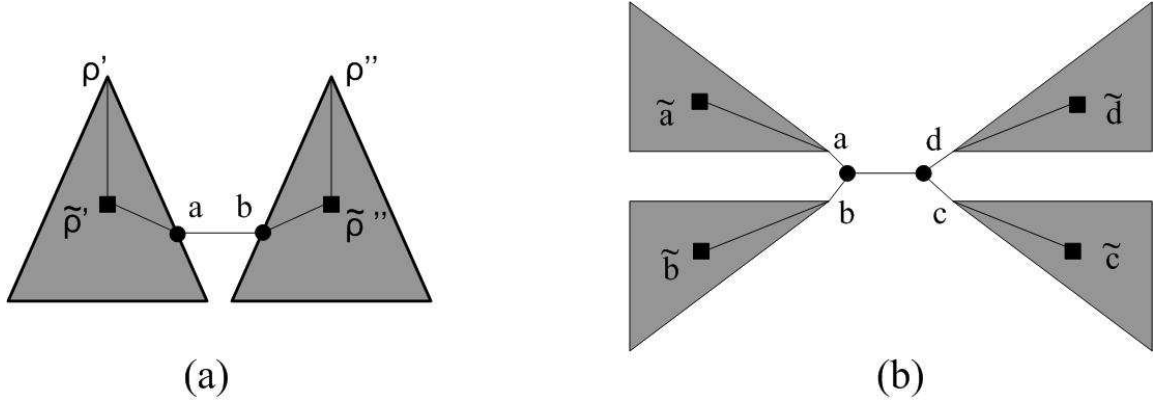

FIGURE 4. Illustration of conditional independence statements in (a) Lemma 5.1 and (b) Lemma 5.2.

Lemma 5.2 gives us a way to reliably estimate lengths of internal paths of $T$, and the subsequent easy corollary shows that if errors in the empirical distances are less than $\epsilon/2$, then the path length estimates are correct to within $\epsilon$.

**Lemma 5.2.** *Let $a_0, b_0, c_0, d_0 \in V(T)$ inducing topology $Q = (a_0, b_0 | c_0, d_0)$ on $T$, and let $l$ be the length of the middle path of $Q$. Let $T_a, T_b, T_c, T_d$ be induced subtrees of $T$ rooted at $a, b, c, d$ respectively, containing $a_0, b_0, c_0$ and $d_0$ respectively, such that $Q, T_a, T_b, T_c, T_d$ are pairwise edge disjoint. Then*

$$FPM(D; \tilde{a}(T_a), \tilde{b}(T_b), \tilde{c}(T_c), \tilde{d}(T_d)) = Q,$$
$$ME(D; \tilde{a}(T_a), \tilde{b}(T_b) | \tilde{c}(T_c), \tilde{d}(T_d)) = l.$$

**Corollary 5.3.** *If $|\hat{D}(x,y) - D(x,y)| < \epsilon/2, \forall x, y \in \{\tilde{a}(T_a), \tilde{b}(T_b), \tilde{c}(T_c), \tilde{d}(T_d)\}$ and $l > \epsilon$, then*

$$FPM(\hat{D}; \tilde{a}(T_a), \tilde{b}(T_b), \tilde{c}(T_c), \tilde{d}(T_d)) = Q,$$
$$|ME(\hat{D}; \tilde{a}(T_a), \tilde{b}(T_b) | \tilde{c}(T_c), \tilde{d}(T_d)) - l| < \epsilon.$$

**Proof of Lemma 5.2:** By repeated application of Lemma 5.1 we obtain

$$D(\tilde{x}(T_x), \tilde{y}(T_y)) = D(x_0, y_0) + D(x_0, \tilde{x}(T_x)) + D(y_0, \tilde{y}(T_y)),$$

for all $x, y \in \{a, b, c, d\}$. Plugging the above equality into the definition of $FPM$ and $ME$ yields the desired result. □

The next Lemma provides a restriction of the triangle inequality for characters and learned characters under the CFN model. As mentioned in the Introduction, the main difficulty with using learned character sequences at internal nodes is that these character sequences depend non-trivially on the leaves of $T$. This destroys the conditional independence relations which turn our distance measures into additive metrics and hinders the identification of speciation events from pairwise distance information. Lemma 5.5 shows a case where the conditional dependence relations induced by using learned character sequences will act in our favor through a version of the triangle inequality: Lemma 5.4.

**Lemma 5.4.** *Let $T'$ be an induced subtree of $T$ rooted at $\rho$ and let $v \in (V(T) \cap T') \setminus \delta(T')$. Then*

$$D(\tilde{\rho}(T'), v) < D(\rho, v) + D(\rho, \tilde{\rho}(T')). \tag{5}$$

**Proof of Lemma 5.4:** See Appendix A. □

Lemma 5.4 provides the foundation for the next result, our main workhorse in the progressive construction of the topology of $T$.

**Lemma 5.5.** *Let $T'$ and $T_d$ be edge disjoint induced subtrees of $T$. Let $o$ be an internal node of $T'$ and let $a, b, c$ be its neighbors in $T'$. Let $T_a, T_b, T_c$ be the clades of $T'$ rooted at $a, b, c$ respectively which do not contain $o$. Suppose that the shortest path from $T_d$ to $o$ does not pass through $b$ or $c$. Then*

$$FPM(D; \tilde{a}(T_a), \tilde{b}(T_b), \tilde{c}(T_c), \tilde{d}(T_d)) = (\tilde{d}, \tilde{a} | \tilde{c}, \tilde{b}).$$

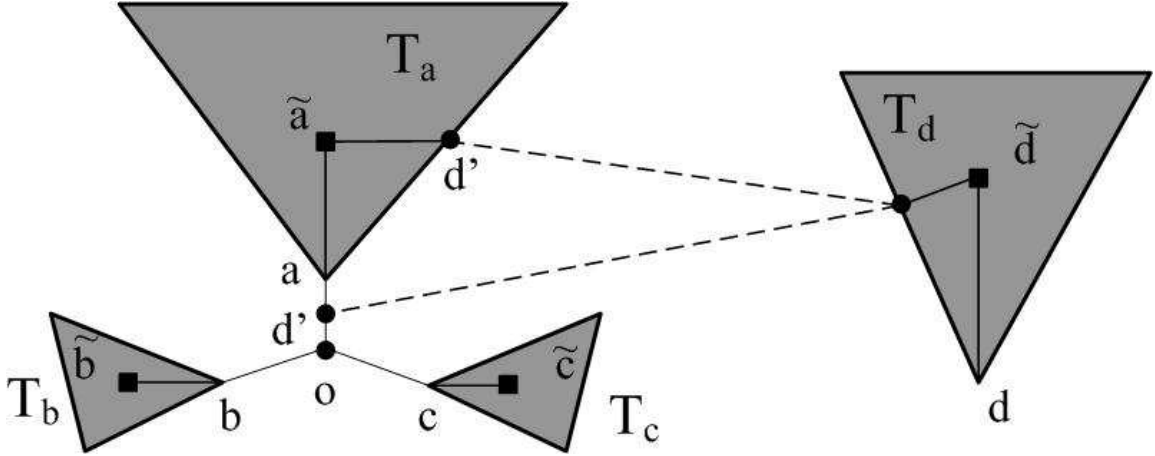

FIGURE 5. Properly connecting induced subtrees by inferring quartets on learned character values.

**Proof of Lemma 5.5:** Our assumptions imply, by repeated application of Lemma 5.1:

$$D(x, y) = D(x, \tilde{x}) + D(x, y) + d(y, \tilde{y}) \quad \text{for all} \quad x, y \in \{a, b, c\} \tag{6}$$

Let $d'$ be the node of $T$ where the path from $T_d$ to $o$ intersects $T_a \cup P(o, a)$. There are two cases: either $d'$ is on the path from $o$ to $a$, or $d' \in T_a \cap V(T)$. See Figure 5.

**Case 1:** $d'$ is on the path from $o$ to $a$. Lemma 5.1 yields:

$$\begin{aligned}
D(\tilde{a}, \tilde{d}) &= D(a, \tilde{a}) + D(a, d') + D(d', \tilde{d}) \\
D(\tilde{b}, \tilde{d}) &= D(b, \tilde{b}) + D(b, o) + D(o, d') + D(d', \tilde{d}') \\
D(\tilde{c}, \tilde{d}) &= D(c, \tilde{c}) + D(c, o) + D(o, d') + D(d', \tilde{d}')
\end{aligned}$$

Combining the above with (6) gives
$$D(\tilde{a},\tilde{b}) + D(\tilde{c},\tilde{d}) = D(\tilde{a},\tilde{c}) + D(\tilde{b},\tilde{d}) = D(\tilde{b},\tilde{c}) + D(\tilde{a},\tilde{d}) + 2D(o,d').$$

**Case 2:** $d' \in T_a$. Lemma 5.1 yields:
$$\begin{aligned} D(\tilde{a},\tilde{d}) &= D(d',\tilde{a}) + D(d',\tilde{d}) \\ D(\tilde{b},\tilde{d}) &= D(b,\tilde{b}) + D(b,o) + D(o,a) + D(a,d') + D(d',\tilde{d}) \\ D(\tilde{c},\tilde{d}) &= D(c,\tilde{c}) + D(c,o) + D(o,a) + D(a,d') + D(d',\tilde{d}) \end{aligned}$$

But by Lemma 5.4, $D(d',\tilde{a}) < D(d',a) + D(a,\tilde{a}')$, and therefore
$$D(\tilde{a},\tilde{b}) + D(\tilde{c},\tilde{d}) = D(\tilde{a},\tilde{c}) + D(\tilde{b},\tilde{d}) \geq D(\tilde{b},\tilde{c}) + D(\tilde{a},\tilde{d}) + 2D(o,a).$$

In both cases the statement of the lemma follows by the definition of FPM. □

**Corollary 5.6.** *Given the hypotheses of Lemma 5.5, suppose $ME(D; a, b|c, d) = l > \epsilon$, and $|\hat{D}(x,y) - D(x,y)| < \epsilon/2$, $\forall x, y \in \{\tilde{a}, \tilde{b}, \tilde{c}, \tilde{d}\}$, then*
$$FPM(\hat{D}; \tilde{a}(T_a), \tilde{b}(T_b), \tilde{c}(T_c), \tilde{d}(T_d)) = (\tilde{d}, \tilde{a}|\tilde{c}, \tilde{b})$$
$$ME(\hat{D}; \tilde{a}_0(T_a), \tilde{b}_0(T_b)|\tilde{c}_0(T_c), \tilde{d}_0(T_d)) > l - \epsilon.$$

In essence, the above corollary states that, when the $(\star)$ condition is obeyed, FPM estimates topological information correctly.

## 6. Implementation details

This section provides all the implementation details for algorithm TREE-MERGE. We let $M, \epsilon, \xi$ be as determined in Section 4. For the remainder of the paper we will assume, unless otherwise stated that condition $(\star)$ holds. As mentioned previously, we maintain an edge-disjoint sub-forest $F$ of $T$, such that, with high probability, the invariants **I1**, **I2** and **I3** are satisfied. Under $(\star)$, we are able to maintain **I1** and **I2** by enforcing **C1** and **C2**.

These invariants are crucial for our ability to ensure that topological information can be reliably estimated from learned sequences (**I1** and **I3**), and that learning of ancestral sequences can be performed reliably (**I2**), by learning via recursive majority on "proper" clades, which are guaranteed to have edge lengths under the phase transition:

**Definition 6.1.** *Given a subtree $T' \in F$, a clade $T''$ of $T'$ is called* proper *if all the edges of $T''$ have estimated lengths shorter than $\lambda_0 - 2\epsilon$.*

Let $v$ be the root of clade $T''$. By $(\star)$ and Lemma 6.2, any proper clade has true edge lengths less than $\lambda_0 - \epsilon$, which guarantees that the learned character sequence $\tilde{v}(T'')$ is at distance at most $\beta$ from the true sequence at $v$. Given an internal node $v$ of some sub-tree $T' \in F$, there are three clades of $T'$ rooted at $v$. Given any edge $e \in E(T')$, we let $T'(v, e)$ denote the unique clade of $T'$ which is rooted at $v$ and does not contain edge $e$. Letting $e_b$ be the long edge described by condition **C2**, we see that for each node

---

Algorithm $d = \text{EdgeLength}(e, T')$.
INPUT: Edge $e$ of a subtree $T' \in F$.
OUTPUT: Estimated length of $e$.
   (1) Let $a, b, c, d \in V(T')$ be the four neighboring nodes of $e$ in $T'$: $e$ is the middle edge of the quartet $Q = (a, b|cd)$.
   (2) **if** some edges of $T'(a, e)$ have not been estimated and/or $T'(a, e)$ is not proper, let $a'$ be the closest descendant of $a$ in $T'(a, e)$ such that $T'(a', e)$ is proper. Set $a = a'$.
   (3) **repeat** the above process for $b, c, d$
   (4) **if** the diameter of $\tilde{Q} = (\tilde{a}(T'(a,e)), \tilde{b}(T'(b,e))|\tilde{c}(T'(c,e)), \tilde{d}(T'(d,e)))$ is higher than $M - \epsilon$ **return FAIL**
   (5) **return** $d = \text{ME}(\hat{D}; \tilde{Q})$.

---

FIGURE 6. Procedure EdgeLength computes the lengths of new edges which may be added to $F$ by joining two of its component trees.

$v \in V(T')$, $T'(v, e_b)$ will be proper. Thus $D(v, \tilde{v}(T'(v, e_b))) < \beta$: under invariant **I2**, we can reliably learn the ancestral sequences of all nodes in $F$. Thus at any point in the algorithm, the sequence at any vertex of $F$ can be learned from some proper clade of $F$, rooted at that vertex.

**Lemma 6.2.** *Suppose the forest $F$ reconstructed by TREE-MERGE at some intermediate step is topologically correct, contains edge-disjoint trees, its edge lengths have been computed to within $\epsilon$ error, obeys conditions **C1** and **C2** (and hence obeys **I1** and **I2**), and all distances between pairs of trees appearing in TreeDistQueue have also been estimated to within $\epsilon$ error. Then, under condition ($\star$), the estimated edge lengths computed at step 4(d) of TREE-MERGE are also correct within $\epsilon$.*

**Proof of Lemma 6.2:** We use the notation of the TREE-MERGE pseudocode. Since $T_1$ and $T_2$ are candidates for being joined, the estimated length of the middle path $(u, v)$ of $Q$, which was computed at a previous iteration of TreeConnection, is at most $2\lambda_0 - 5\epsilon$, and is correct up to $\epsilon$ by our hypothesis. Similarly, $(a, b)$ and $(c, d)$ are edges of $F$ and obey **I1**. Thus all edges of $Q$ are less than $2\lambda_0 - 4\epsilon$.

Since the length of $(u, v)$ has been estimated, we only need to estimate edges $(a, u)$, $(b, u)$, $(c, v)$ and $(d, v)$. It is an easy exercise to prove that the two neighbors of $a$ either root a proper clade in $T_1$ which does not contain $a$, or have a neighbor who roots such a clade. This follows by **C2**. It follows that the edge $(a, u)$ can be estimated from a quartet of diameter at most $6\lambda_0 + 2\beta$. Thus the procedure EdgeLength will estimate it within $\epsilon$. We proceed by symmetry for the other edges. □

We next give the details of the TreeConnection and TreeDistance subroutines, which find the topologically correct way to connect two sub-trees $T_1, T_2$. TreeConnection requires seed nodes $u_i \in V(T_i)$, $i \in \{1, 2\}$, rooting proper clades $T'_1$ of $T_1$ and $T'_2$ of $T_2$ such that $\hat{D}(\tilde{u}_1(T_1'), \tilde{u}_1(T_1')) < M/3 - \epsilon$. The algorithm proceeds by moving along $T_1$ and $T_2$ in the direction indicated by quartet tests around the current candidate node. Lemma 6.3 shows that TreeConnection will find the correct link between $T_1$ and $T_2$, given "sufficiently close" seed nodes.

**Lemma 6.3.** *Suppose Condition ($\star$) holds. Let $T_1$ and $T_2$ be subtrees of $T$ satisfying invariants **I1** and **I2**, and let $P = (v_1, v_2)$ be the path joining them in $T$, with $v_i \in V(T) \cap e_i$, and $e_i \in E(T_i)$, $i = 1, 2$. Let $E_1 \subset E(T_1)$ and $E_2 \subset E(T_2)$, such that the following hold:*

- *$e_1 \in E_1$ and $e_2 \in E_2$*
- *there exist proper clades $T_1'$ and $T_2'$ of $T_1$ and $T_2$, rooted at $u_1 \in V(E_1)$ and $u_2 \in V(E_2)$ respectively, such that $D(\tilde{u}_1(T_1'), \tilde{u}_2(T_2')) < M/3$.*

*Then $e_1 \in C_1$ and $e_2 \in C_2$, where $(C_1, C_2) = \text{TreeConnection}(T_1, T_2, E_1, E_2)$. Note that $C_i$ either contains a single edge or three adjacent edges of $T_i$. If all the edges of $T'$ have length at least $2\epsilon$, then $C_1 = \{e_1\}$ and $C_2 = \{e_2\}$. Furthermore, if $|C_i| = 3$, then $D(v_i, c_i) < 2\epsilon$, where $c_i$ is the center node of $C_i$.*

**Proof of Lemma 6.3:** By Lemma 5.1,

$$D(\tilde{u}_1(T_1'), \tilde{u}_2(T_2')) = D(\tilde{u}_1(T_1'), v_1) + D(v_1, \tilde{u}_2(T_2')).$$

Therefore $D(v_1, \tilde{u}_2(T_2')) < M/3$. Now suppose $e_1 = (v_1', v_1'')$. This edge defines two clades in $T_1$, at least one of which is proper; we may assume w.l.o.g. that $v_1'$ roots a proper clade. Let $u_1', u_1''$ be the descendants of $v_1'$ in said clade, and let $T_1', T_2'$ be the corresponding sub-clades. Then

$$D(\tilde{u}_1''(T_1''), \tilde{u}_2(T_2')) = D(\tilde{u}_1''(T_1''), u_1'') + D(u_1'', v_1) + D(v_1, \tilde{u}_2(T_2')) < \beta + 4\lambda_0 + \frac{M}{3} < \frac{M}{2},$$

thus $u_1 = v_1', u_1', u_1''$ will satisfy the conditions of step 1 in subroutine TreeConnection.

In turn, let $u_1, u_1', u_1''$ satisfy the conditions of step 1 in subroutine TreeConnection. Then at least one of $u_1', u_1''$ is not on the path $P(v_1, u_1)$. We may assume $u_1' \notin P(v_1, u_1)$; then $u_1 \in P(v_1, u_1')$ and thus

$$M/2 + \epsilon > \hat{D}(\tilde{u}_1'(T_1'), \tilde{u}_2(T_2')) = D(\tilde{u}_1'(T_1'), u_1) + D(u_1, \tilde{u}_2(T_2')).$$

Thus $M/2 + \epsilon > D(u_1, \tilde{u}_2(T_2')) = D(u_1, v_1) + D(v_1, \tilde{u}_2(T_2'))$.

Letting $a, b, c$ be as described in TreeConnection, Lemmas 5.1 and 5.4 imply that

$$D(\tilde{a}(T_1), \tilde{u}_2(T_2')), D(\tilde{b}(T_1), \tilde{u}_2(T_2')), D(\tilde{c}(T_1), \tilde{u}_2(T_2')) < M/2 + \epsilon + 3\lambda_0 + \beta < M - 2\lambda_0.$$

Suppose w.l.o.g. that $(a, b|c, u_2)$ is the true quartet topology induced by $T$. Suppose the middle edge of $(a, b|c, u_2)$, namely $(u_1, v_1)$, is shorter than $\epsilon$. Since $T_1$ obeys invariant (1), this implies that $v_1$ is indeed a neighbor of $u_1$ in $T'$. Furthermore,

$$ME(D; (\tilde{a}(T_a), \tilde{c}(T_c)|\tilde{b}(T_b), \tilde{u}_2(T_2'))) < 0 \Rightarrow \text{ME}(\hat{D}; (\tilde{a}(T_a), \tilde{c}(T_c)|\tilde{b}(T_b), \tilde{u}_2(T_2'))) < \epsilon.$$

Similarly $ME(\hat{D}; (\tilde{b}(T_b), \tilde{c}(T_c)|\tilde{a}(T_a), \tilde{u}_2(T_2'))) < \epsilon$. So in case TreeConnection picks the wrong direction, it will immediately exit correctly at step 2.f.

---

Algorithm $(E_1, E_2) = \text{TreeConnection}(T_1, T_2, E_1, E_2)$
INPUT: Subtrees $T_1, T_2 \in F$ and candidate sets $E_1 \subset E(T_1), E_2 \subset E(T_2)$ containing the endpoints of the path joining $T_1$ and $T_2$.
OUTPUT: Refined candidate sets $E_1 \in E(T_1), E_2 \in E(T_2)$.
  (1) Let $u_1 \in V(E_1)$ and $u_2 \in V(E_2)$, and $T_2'$ be a proper clade of $T_2$ rooted at $u_2$. Let $T_1', T_1''$ be edge disjoint proper subtrees of $T_1$ rooted at neighbors of $u_1$ in $T_1$, $u_1'$ and $u_1''$. Suppose $\hat{D}(\tilde{u}_1'(T_1'), \tilde{u}_2(T_2')) < M/2 + \epsilon$, and $\hat{D}(\tilde{u}_1''(T_1''), \tilde{u}_2(T_2')) < M/2 + \epsilon$. If no such $u_1, u_2$ exist, then **return** $\emptyset$.
  (2) **while** $|E_1| > 1$
     (a) Let $a, b, c$ be the neighbors of $u_1$.
     (b) Let $T_a, T_b, T_c$ be edge disjoint clades of $T_1$ rooted at $a, b, c$ respectively.
     (c) **if** $T_a$ is not proper
        • Let $a'$ be the descendant of $a$ in $T_a$ which roots a maximal sub-clade of $T_a$ which is proper (does not contain the long edge of $T_a$. Set $a = a'$, $T_a = T_{a'}$.
     (d) Do the same as above for $b, c$.
     (e) Let $Q = \text{FPM}(\hat{D}; \tilde{a}(T_a), \tilde{b}(T_b), \tilde{c}(T_c), \tilde{u}_2(T_2'))$, with $Q = (\tilde{u}_2, x|y, z)$, $\{x, y, z\} = \{a, b, c\}$.
     (f) If $\text{ME}(\hat{D}; Q) < \epsilon$, set $E_1$ to the set of edges incident to $u_1$ and **go to** step (3).
     (g) Set $E_1 = E_1 \cap (E(T_x) \cup \{(u_1, x)\})$.
     (h) Set $u_1 = x$.
  (3) Repeat the same process to restrict $E_2 \subset E(T_2)$.

FIGURE 7. Subroutine $\text{TreeConnection}(T_1, T_2, E_1, E_2)$ finds the edges of $T_1, T_2$ containing the endpoints of their connecting path $P$. TreeConnection will output a single edge per tree, or, in case $P$ connects too close to an existing node, the edges adjacent to that node.

Alternatively, if $D(u_1, v_1) > \epsilon$, TreeConnection will pick the correct "direction", by Lemma 5.5 and Corollary 5.6. Furthermore, if the middle edge is longer than $2\epsilon$, its estimated length will be also larger than $\epsilon$, and thus the algorithm will proceed to the next iteration of the while loop. This implies the last statement of our lemma. Also, if $v_1$ does not lie on the path $P(u_1, c)$, then $D(u_1, v_1) > D(u_1, c) > 2\epsilon$, and thus the algorithm will proceed to the next iteration of the while loop.

To complete the argument, either: $v_1 \in P(u_1, c)$ or $c \in P(v_1, u_1)$. In the first case

$$D(c, \tilde{u}_2(T_2')) < D(u_1, \tilde{u}_2(T_2')) + 2\lambda_0 < M/2 + \epsilon + 2\lambda_0,$$

and the procedure will terminate at the next iteration of the while loop. In the latter case,

$$D(c, \tilde{u}_2(T_2')) < D(u_1, \tilde{u}_2(T_2')) < M/2 + \epsilon,$$

---

Subroutine $d = \text{TreeDistance}(T_1, T_2, E_1, E_2)$
INPUT: Subtrees $T_1, T_2 \in F$ and candidate sets $E_1 \subset E(T_1), E_2 \subset E(T_2)$ containing the endpoints of the path $P$ joining $T_1$ and $T_2$. OUTPUT: Estimated length of $P$.
  (1) **if** $|E_1| = |E_2| = 1$
      - Let $Q = (a, b|c, d)$ with $E_1 = \{(a,b)\}, E_2 = \{(c,d)\}$. Let $e$ be the middle edge of $Q$.
      - Let $T'$ be the tree given by joining $T_1$ and $T_2$ according to $Q$.
      - **return** $\text{EdgeLength}(e, T')$.
  (2) **else**
      - Let $E_1 = \{(v, v_1), (v, v_2), (v, v_3)\}$.
      - **return** $\min\{\text{TreeDistance}(T_1(v, (v, v_i)), T_2, \{(j, k)\}, E_2), \{i, j, k\} = \{1, 2, 3\}\}$

---

FIGURE 8. Subroutine $\text{TreeDistance}(T_1, T_2, E_1, E_2)$ estimates the length of the path connecting $T_1, T_2$, based on the set of possible connection edges $E_1, E_2$, output by $\text{TreeConnection}(T_1, T_2)$.

and we can proceed by induction on $|E_1|$ to show that at every step TreeConnection picks the correct direction or exits correctly. □

**Lemma 6.4.** *Assume all the hypotheses and notation of Lemma 6.3 hold. Let $(C_1, C_2) = TreeConnection(T_1, T_2, E_1, E_2)$. Then $|TreeDistance(T_1, T_2, C_1, C_2) - L(P)| < \epsilon$.*

**Proof of Lemma 6.4:** All the quartets inspected by TreeDistance where previously inspected by TreeConnection as well. The proof of Lemma 6.3 shows that the true diameters of all said quartets are less than $M - \epsilon$. Thus $(\star)$ and Corollary 5.6 imply that all path lengths returned by EdgeLength will be $\epsilon$-approximations of the true path lengths. Furthermore, it is an easy exercise to see that TreeDistance returns $L(P)$ in the event that $\hat{D} = D$. The conclusion follows trivially. □

## 7. Correctness, stopping criteria and running time analysis

In this section we prove the three main theorems of the paper, which were stated in Section 4.

**Theorem 4.4:** *If $(\star)$ holds, algorithm TREE-MERGE returns a topologically correct sub-forest $F$ of $T$ satisfying invariants **I1**, **I2**, **I3**.*

**Proof of Theorem 4.4:** We proceed inductively to show that invariants **I1**,**I2**,**I3** are always obeyed by TREE-MERGE, and additionally that:

  **C1':** edge lengths computed by EdgeLength are correct within $\epsilon$.
  **C2':** all connections computed by TreeConnection are correct,
  **C3':** tree distances contained or previously contained in TreeDistQueue are correct within $\epsilon$.

All of the above hold trivially under $(\star)$ in the case of $F = X$, which is our base case. By Lemma 6.2, **C1**$'$ will hold inductively. Then steps 4.c, 4.e and 4.f, together with Lemma 6.3, show that **C1** and **C2** are never violated. In turn, **C1**, **C2** and **C1**$'$ imply that invariants **I1**, **I2** hold for the next iteration of the algorithm.

Lemma 6.3 shows that the TreeConnection sub-routine returns a topologically correct way of connecting to components of $F$, as long as **I1** and **I2** hold and the candidate edge sets $E_1$ and $E_2$ contain the endpoints of the correct linking path. It is an easy argument to show that if the previously computed connections were correct, then step UpdateTreeDistQueue calls TreeConnection with appropriate sets of candidate edges. This establishes condition **C2**$'$. Lemma 6.4 together with condition **C2**$'$ thus imply condition **C3**$'$.

To complete the proof of the theorem, it remains to show that invariant **I3** is obeyed, namely the components of $F$ are disjoint in $T$. Suppose by contradiction that $T_1, T_2$ are the first pair of subtrees that are joined such that the path $P$ linking them overlaps subtree $T_3 \in F$. Suppose $l$ is the true length and $l'$ the estimated length of the path $P$, computed by TreeDistance$(T_1, T_2)$. Let the distance between $T_1$ and $T_3$ be $l_1$, the distance between $T_2$ and $T_3$ be $l_2$, and let $l'_1$ and $l'_2$ be the corresponding estimated distances. Since $T_1, T_2$ were joined, $3\epsilon < l' < 2\lambda_0 - 5\epsilon$. Therefore $l < l' + \epsilon < 2\lambda_0 - 4\epsilon$. Then $l_1 + l_2 < l < 2\lambda_0$, which implies that the two estimated distances, $l'_1$ and $l'_2$, were computed at a previous step of TREE-MERGE. Then

$$l'_1 + l'_2 - 2\epsilon < l_1 + l_2 < l < l' + \epsilon \Rightarrow l'_1 + l'_2 < l' + 3\epsilon,$$

which contradicts step 4.d of TREE-MERGE. $\square$

**Theorem 4.5:** *Let $T$ satisfy $6\epsilon \leq L(e) \leq \lambda_0 - 3\epsilon, \forall e \in E(T)$. Then given $N$ independent samples $\chi_1, \ldots, \chi_N$ from the character distribution $\mathbb{P}_{T,L}$, $T$ will be fully and correctly recovered by TREE-MERGE with probability at least $1 - \xi$.*

**Proof of Theorem 4.5:** As before, $(\star)$ holds with probability $1 - \xi$. Theorem 4.4 shows that the output of the algorithm is topologically correct. It remains to prove that, under the additional hypotheses of the present theorem, TREE-MERGE will not terminate before the full topology is resolved.

Let us suppose that TREE-MERGE outputs a forest $F$ with more than one component. Let $T_F$ be the tree given by collapsing every connected component of $F$ into a single node. All edges of $T_F$ correspond to single edges of $T$. Since all edges in $T$ are longer than $6\epsilon$, Lemma 6.3 shows that TreeConnection will always output well-defined connections (i.e. candidate edge sets of cardinality 1), and moreover all internal edge estimates will be longer than $5\epsilon$. Thus condition **C1** is never violated and TREE-MERGE will never reject a candidate pair at steps 4.c. or 4.e.

Let $T_1$ and $T_2$ form a cherry of $T_F$. Suppose the common neighbor of $T_1$ and $T_2$ in $T_F$ does not correspond to another component of $F$. The length of the path $P$ joining $T_1$ and $T_2$ will be less than $2\lambda_0 - 6\epsilon$ and therefore the pair $T_1, T_2$ was inserted into TreeDistQueue. Letting $T_{new}$ be the tree formed by joining $T_1$ and $T_2$, we see that all edges in $T_{new}$ other than the one corresponding to $P$, correspond to single edges of $T$, and have lengths between $6\epsilon$ and $\lambda_0 - 3\epsilon$. Thus $T_{new}$ will satisfy condition **C2** under $(\star)$, and will not get rejected at step 4.f. of TREE-MERGE.

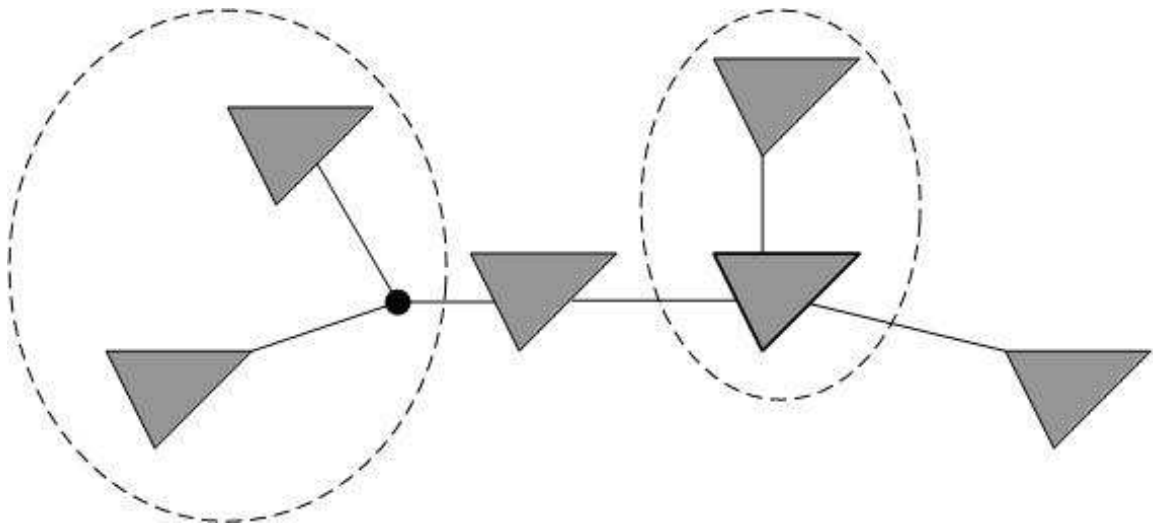

FIGURE 9. The connectivity graph induced by $T$ on $F$, with configurations of pairs of subtrees which can be joined by TREE-MERGE.

Alternatively, suppose the common neighbor of $T_1$ and $T_2$ in $T_F$ corresponds to $T_3 \in F$. Then $T_3$ contains at most one "long" edge, by **C2**. Thus the tree obtained by joining $T_3$ to $T_1$ will also contain at most one long edge, as all edges of the new tree which are not also edges of $T_3$ correspond to single edges of $T$, and thus are "short". Thus, again, the pair $T_1, T_3$ will not be rejected at step 4.f.

The only remaining possibility is that the candidate pair $T_1, T_2$ gets rejected at step 4.g.. From our selection of the candidate pair, we can see that the joined tree $T_{new}$ is in fact edge disjoint from all other trees in $F$. Let $T_k$ be the tree causing the rejection at step 4.d.. We let $l, l_1, l_2$ be the lengths of the paths joining $T_1, T_2$, $T_1, T_k$ and $T_2, T_k$, and $l', l'_1, l'_2$ be the corresponding estimated tree distances. Since $T_{new}$ and $T_k$ are edge disjoint and all edges of $T$ have length at least $6\epsilon$, it is a simple argument to show that $l_1 + l_2 > l + 6\epsilon$, and thus $l'_1 + l'_2 < l' + 3\epsilon$ cannot occur under $(\star)$. □

**Theorem 4.6:** *TREE-MERGE always terminates in $O(Nn^2 + n^3)$ time, where the proportionality constant is a* decreasing *function of $\xi$ and $\epsilon$.*

**Proof of Theorem 4.6:** Steps 1-3 of TREE-MERGE trivially take $O(n^2 N + n^2 \log(n))$ time. Every iteration of Step 4, the main loop of the algorithm, either reduces the size of the forest $F$ by one, or determines that a pair of trees in $F$ cannot be merged. Each time the forest gets modified, a single new tree is produced. Since the forest is modified at most $n - 1$ times, throughout the life of the algorithm there are at most $n^2$ tree pairs being inserted/popped from TreeDistQueue. Thus there are at most $O(n^2)$ iterations of step 4. In particular, the total time spent in steps 4.a-d is $O(n^2 \log(n))$.

In order to verify **C1** and **C2** on $T_{new}$, one only needs to compute a fixed number of edge lengths in addition to the ones already in the forest, so again, steps 4.e-f take $O(n^2)$ time. The verification at step 4.g takes linear time per iteration, so at most $O(n^3)$ time.

Step 4.h is equivalent to a modification of the forest, so it only occurs at most $n-1$ times. Learning the sequences of new proper clades can be done in a bottom-up fashion, such that each new root can be computed through a single recursive majority step. Thus step 4.h.(ii) takes time proportional to the number of new learned sequences, and its contribution to the total running time is $O(nN)$. Similarly step 4h.(iii) will contribute at most $O(n^2N + n^2 \log(n))$.

Now suppose that $|V(T_{new})| = t$ and $|F| = s$. The subroutine TreeConnection runs in time at most linear in the sizes of its input subtrees. Thus one iteration of UpdateTreeDistQueue will spend $O(st + (n-t))$ time for building the new tree connections, and $O(s \log(n))$ time in the insertion and deletion of tree pairs from TreeDistQueue. Summing over all iterations, the total time spent in UpdateTreeDistQueue is $O(n^3)$. □

## 8. Final remarks

A simple amortized argument shows that step 4.h, as detailed here, only takes $O(n^2(N + \log(n)))$ time. We do not include this argument for the sake of brevity. Thus the verification of step 4.g is the true running-time bottleneck of the algorithm. In practice, this verification should be rendered somewhat redundant by the fact that at every step we join the pair of trees which are closest, but sadly this is not sufficient for a formal proof of correctness.

Our methods here are general enough to specify a all-purpose phylogeny reconstruction algorithm. The bounds we require on edge lengths can in fact be relaxed at the cost of longer sequence lengths. If one is able to estimate the lengths of the already-constructed edges, then one can also estimate the expected disagreement between the real and learned sequences at internal nodes. Similarly, with given sequence lengths one can infer a "robust" area of the space given by $M$ and $\epsilon$: for every estimated distance we can infer an $\epsilon$ that is larger than the estimation error with high probability. If the phylogeny can be progressively disambiguated from the available distances, given their expected errors, then we have achieved our purpose.

Indeed, these ideas are at the base of the methods in [19]. The availability of Mossel's techniques for inferring ancestral sequences simply give us a very powerful tool for "reaching deeper" into the phylogenetic tree and improving on classical distance methods without departing too much from their simplicity.

## Acknowledgment

We thank Elchanan Mossel for invaluable discussions regarding reconstruction of ancestral sequences. Thanks to Costis Daskalakis, Elchanan Mossel and Sebastien Roch for pointing out errors in a preliminary version of the paper. Radu Mihaescu was supported by a National Science Foundation Graduate Fellowship, by the Fannie an John

Hertz Foundation graduate fellowship and by the CIPRES project. All other authors were supported by CIPRES.

Appendix A. The triangle inequality for the CFN model

The main purpose of this Appendix is to provide a proof of Lemma 5.4:

**Lemma 5.4.** *Let $T'$ be an induced subtree of $T$ rooted at $\rho$ and let $v \in (V(T) \cap T') \setminus \delta(T')$. Then*
$$D(\tilde{\rho}(T'), v) < D(\rho, v) + D(\rho, \tilde{\rho}(T')).$$
*with $\tilde{\rho}(T')$ a "learned" character value, where the learning occurs by any bottom-up recursive majority algorithm on $T'$, as outlined in Section 3.*

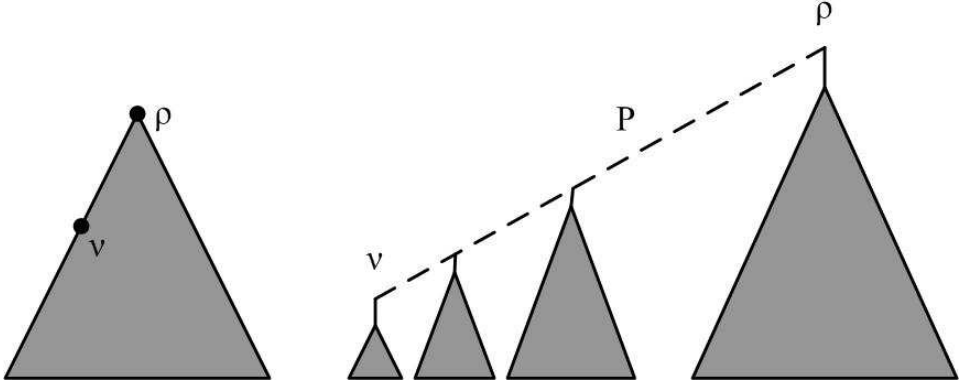

Figure 10. Tree configuration for the proof of Lemma 5.4.

We begin by introducing an alternative representation of the CFN model under a percolation framework. This intuitive view lies at the root of the theoretical results regarding information flow on trees in [16] and [17].

Let $p(e) < 0.5$ be the probabilities of mutation along edges $e \in E(V)$ for a CFN model on $T$. Let $\alpha(e)$ be independent random variables such that

$$\alpha(e) = \begin{cases} 1 \text{ with probability } 1 - 2p(e) \\ 0 \text{ with probability } 2p(e). \end{cases}$$

Suppose each edge $e$ in $T$ carries a survival probability $\theta(e) = 1 - 2p(e)$, such that the edge $e$ is deleted if $\alpha(e) = 0$. After removing the destroyed edges, each surviving connected component $C$ receives a single character value $\chi(C)$, by tossing an independent unbiased coin. We write $u \leftrightarrow v$ for the event that the two nodes $u, v$ are in the same connected component and $C_v$ for the component containing $v$.

It is easy to see that the joint probability distribution on character values at $V(T)$ produced under this alternative model, $\mathbb{P}_{T,\theta}$, is the same as the one induced by the original CFN model: $\mathbb{P}_{T,L}$, where $L(e) = -\log(1 - 2p(e))/2 = -\log(\theta(e))/2$. As before, $D$ is the distance between uniform binary random variables defined in Section 2.

**Proof or Lemma 5.4:** Let $\tilde{\rho}$ denote $\tilde{\rho}(T')$ for brevity. The lemma is equivalent to $\mathbb{E}[v\tilde{\rho}] \geq \mathbb{E}[\rho v]\mathbb{E}[\rho\tilde{\rho}]$. It is an easy exercise to show that, in turn, this is equivalent to $\mathbb{P}[\rho = \tilde{\rho}|\rho = v] \geq \mathbb{P}[\rho = \tilde{\rho}|\rho \neq v]$. By symmetry, we may assume for the rest of the proof that $\rho = 1$, so $\chi(C_\rho) = 1$, and our task reduces to showing that

$$\mathbb{P}[\tilde{\rho} = 1|\rho = v = 1] \geq \mathbb{P}[\tilde{\rho} = 1|\rho = 1 \neq v].$$

Let $E(P) = \{e_1, \ldots, e_s\}$ denote the edges of the path $P = P(\rho, v)$, and let $V(P) = \{v_1, \ldots, v_s = v\}$ be the nodes of $P$, other than $\rho$. Then $\mathbf{1}(\rho \leftrightarrow v) = \mathbf{1}(\alpha(E(P)) \equiv 1)$. We proceed by way of a standard coupling argument.

Suppose $\alpha(E(T))$ is such that such that $\rho \not\leftrightarrow v$. Given a set of values $\chi_0$ for the characters $\chi(C), C \neq C_v, C \neq C_\rho$,

$$\mathbb{P}[\chi(C_v) = 1, \chi(C_{\neq v,\rho}) = \chi_0|\alpha] = \mathbb{P}[\chi(C_v) = -1, \chi(C_{\neq v,\rho}) = \chi_0|\alpha].$$

Now $\tilde{\rho}$ is a recursive majority function in the character values at $\delta T'$, and is therefore coordinate-wise increasing in the values of those characters. Moreover $\chi(\delta T')$ in the event $\mathbf{1}[\chi(C_v) = 1, \chi(C_\rho) = 1, \chi(C_{\neq v,\rho}) = \chi_0]$ is coordinate-wise larger than $\chi(\delta T')$ in the event $\mathbf{1}[\chi(C_v) = -1, \chi(C_\rho) = 1, \chi(C_{\neq v,\rho}) = \chi_0]$, while the probabilities of the two events, conditioned on the values $\alpha$, are the same. Summing over all values $\chi_0$ and all values $\alpha$ such that $\rho \not\leftrightarrow v$,

$$\mathbb{P}[\tilde{\rho} = 1|\rho \not\leftrightarrow v, v = 1] \geq \mathbb{P}[\tilde{\rho} = 1|\rho \not\leftrightarrow v, v = -1] = \mathbb{P}[\tilde{\rho} = 1|v = -1]. \tag{7}$$

For any $x \in \{\pm 1\}^s$ and any $b \in \{\pm 1\}^t$ with $t = |E(T)| - s$, an identical argument to the one above shows that

$$\mathbb{P}[\tilde{\rho} = 1|\alpha(E(T)\setminus E(P)) = b, \chi(V(P)) \equiv 1] \geq \mathbb{P}[\tilde{\rho} = 1|\alpha(E(T)\setminus E(P)) = b, \chi(V(P)) = x].$$

We observe that $\mathbf{1}[\rho \leftrightarrow v] = \mathbf{1}[\alpha(E(P)) \equiv 1]$ implies $\mathbf{1}[\chi(V(P)) \equiv 1]$ and that $\alpha(E(T)\setminus E(P))$ and $\alpha(E(P))$ are independent, thus $\alpha(E(T)\setminus E(P))$ and $\chi(V(P))$ are independent. Therefore

$$\begin{aligned}
\mathbb{P}[\tilde{\rho} = 1|\alpha(E(T) \setminus E(P)) = b, v \leftrightarrow \rho] &\geq \mathbb{P}[\tilde{\rho} = 1|\alpha(E(T) \setminus E(P)) = b, \chi(V(P)) = x], \forall x \\
\Rightarrow \mathbb{P}[\tilde{\rho} = 1|v \leftrightarrow \rho] &\geq \mathbb{P}[\tilde{\rho} = 1|\chi(V(P)) = x], \forall x \\
\Rightarrow \mathbb{P}[\tilde{\rho} = 1|v \leftrightarrow \rho] &\geq \mathbb{P}[\tilde{\rho} = 1|\alpha(E(P)) = a, v = 1], \forall a \in \{\pm 1\}^s, a \not\equiv 1 \\
\Rightarrow \mathbb{P}[\tilde{\rho} = 1|v \leftrightarrow \rho] &\geq \mathbb{P}[\tilde{\rho} = 1|v \not\leftrightarrow \rho, v = 1].
\end{aligned}$$

The first implication follows from summation over all values of $b$. The second comes from summation over all values of $x$ such that $x_s = v = 1$ and $x$ is compatible with $\alpha(E(P)) = a$. The third implication follows from summing over all values $a \in \{\pm 1\}^s, a \not\equiv 1$.

The last inequality, together with (7), implies

$$\mathbb{P}[\tilde{\rho} = 1|v = 1] \geq \mathbb{P}[\tilde{\rho} = 1|v \not\leftrightarrow \rho, v = 1] \geq \mathbb{P}[\tilde{\rho} = 1|v \not\leftrightarrow \rho, v = -1] = \mathbb{P}[\tilde{\rho} = 1|v = -1].$$

□

Appendix B. Applicability to other molecular models of evolution

Our method implies similar results for all group based models of evolution, where character alphabet is a group $G$ admitting a non-trivial morphism $\phi : G \to \mathbb{Z}_2$. In a group-based model of evolution, the probability of transformation of the character $\chi$ from state $a$ to state $b$ along any edge $e$ of the tree only depends on $a^{-1}b$. In other words, for an edge $e = (u, v) \in E(T)$,

$$\mathbb{P}(\chi(v) = b | \chi(u) = a) = p_e(a^{-1}b.$$

By the definition of a morphism,

$$\phi(a) \neq \phi(b) \quad \Leftrightarrow \quad \phi(a^{-1}b) = -1.$$

Thus

$$\mathbb{P}[\phi(\chi(u)) \neq \phi(\chi(v)) | \chi(u) = a] = \mathbb{P}[\phi(\chi(u)) \neq \phi(\chi(v)) | \phi(\chi(u)) = \phi(a)]$$
$$= \sum_{g \in \phi^{-1}(-1)} p(g),$$

which does not depend on $a$ and implicitly does not depend on $\phi(a)$.

We can then reduce any such model to the binary one by identifying a state $g \in G$ to $\phi(g) \in \mathbb{Z}_2$ and applying our analysis mutatis mutandis. The most notable example of group based model of evolution satisfying our requirements is the Kimura 3ST model [15], which is realized by the group $\mathbb{Z}_2 \times \mathbb{Z}_2$ [20]. We also note that Kimura 3ST is a generalization of the well known Jukes-Cantor model.

# References

[1] N. Abe and N. K. Warmuth, On the Computational Complexity of Approximating Probability Distributions by Probabilistic Automata, Machine Learning, 9(2–3), 205–260, 1992.

[2] N. Alon and J. Spencer. The Probabilistic Method. Wiley, 1992.

[3] J. Cavender, Taxonomy with confidence, Mathematical Biosciences, 40, 271–280, 1978.

[4] M. Cryan, L. A. Goldberg, and P. W. Goldberg, Evolutionary Trees can be Learned in Polynomial Time in the Two-State General Markov Model, SIAM Journal on Computing, 31(2), 375–397, 2002.

[5] C. Daskalakis, C. Hill, A. Jaffe, R. Mihaescu, E. Mossel, S. Rao, Maximal Accurate Forest from Distance Matrices, RECOMB 2006.

[6] C. Daskalakis, E. Mossel, S. Roche, Optimal Phylogenetic Reconstruction, 38th ACM Symposium on Theory of Computing, STOC 2006

[7] C. Daskalakis, E. Mossel, S. Roche, Phylogenies without Branch Bounds: Contracting the Short, Pruning the Deep , http://arxiv.org/abs/0712.0171.

[8] P. L. Erdos, M. Steel, L. Szekely, and T. Warnow, A Few Logs Suffice to Build (Almost) all Trees (I), Random Structures and Algorithms, 14, 153–184, 1997.

[9] P. L. Erdos, M. A. Steel, L. A. Szekely, and T. J. Warnow, A Few Logs Suffice to Build (Almost) all Trees (II), Theoretical Computer Science, 221(1–2), 77–118, 1999.

[10] J. S. Farris, A probability model for inferring evolutionary trees, *Systematic Zoology,* vol. 22, pp. 250-256, 1973.

[11] J. Felsenstein. *Inferring Phylogenies*. Sinauer, New York, New York, 2004.

[12] Fischer, M. and Steel, M. Sequence length bounds for resolving a deep phylogenetic divergence. Submitted to Journal of Theoretical Biology.

[13] Ilan Gronau, Shlomo Moran, and Sagi Snir, Fast and reliable reconstruction of phylogenetic trees with very short edges. To appear in SODA, 2008.

[14] T. E. Harris, A correlation inequality for Markov processes in partially ordered state spaces, Ann. Probab, 5, 451–454, 1977.

[15] M. Kimura, Estimation of evolutionary distances between homologous nucleotide sequences, PNAS January 1, 1981 vol. 78 no. 1 454-458.

[16] E. Mossel, On the impossibility of reconstructing ancestral data and phylogenies, 669–678 (2003) Jour. Comput. Bio. 10 no.5

[17] E. Mossel, Phase transitions in phylogeny, Transactions of the American Mathematical Society, Vol. 356 No. 6, 2379–2404 (electronic) (2004).

[18] E Mossel, Distorted metrics on trees and phylogenetic forests, IEEE/ACM Trans. Comput. Biology Bioinform. 4(1): 108-116, 2007.

[19] S Roch, Sequence Length Requirement of Distance-Based Phylogeny Reconstruction: Breaking the Polynomial Barrier, Proceedings of the 49th IEEE Symposium on Foundations of Computer Science, FOCS 2008.

[20] C. Semple and M. Steel. Phylogenetics, volume 22 of *Mathematics and its Applications series*.

[21] M. Steel. My Favourite Conjecture. Preprint, 2001.